\documentclass[%
 reprint,
 amsmath,amssymb,
 aps,
prl,
floatfix,
]{revtex4-1}

\usepackage{graphicx}
\usepackage{dcolumn}
\usepackage{bm}

\usepackage{hyperref}
\usepackage{siunitx}
\usepackage{ucs}
\usepackage[utf8]{inputenc}
\usepackage{balance}
\usepackage{fontenc}
\newcommand{\MS}[1]{\color{black}{#1}\color{black}}
\newcommand{\ES}[1]{\color{black}{#1} \color{black}}

\begin{document}


\title{Nonmonotonic temperature dependence of the thermopower\\of atomic-size gold contacts}

\author{Thomas B. M\"oller, Marcel Strohmeier, Johannes Boneberg, Paul Leiderer, Wolfgang Belzig, and Elke Scheer}
\email{elke.scheer@uni-konstanz.de}

\affiliation{Department of Physics, University of Konstanz, 78457 Konstanz, Germany}

\date{\today}

\begin{abstract}
\vspace{0.3cm}
\noindent We report measurements of the thermopower of
atomic-size gold contacts realized by the mechanically controllable break junction (MCBJ) technique over a temperature range from \SIrange{18}{295}{\kelvin}. A thermometer included in the lithographic structure close to the constriction provides a direct measurement of the temperature increase generated by heating
one side of the contact with a focused laser beam. While the conductance histograms confirm the quantum nature of the transport, we observe a nonmonotonic temperature dependence of the ensemble-averaged thermopower with a minimum
of \SI{-2}{\micro\volt\per\kelvin} at about \SI{150}{\kelvin}. \ES{The values for the thermopower obtained} at the lowest and the high temperature are compatible with values reported in the literature, but the nonmonotonic behavior in between disagrees with the expected linear dependence for quantum coherent conductors described by the Landauer formula. \ES{We develop a theoretical model based on an energy dependent transmission function that qualitatively } \MS{reproduces } \ES{the  nonmonotonic behavior, but fails qualitatively.} We therefore interpret our data as a result of phonon contributions to the thermopower \ES{beyond the Landauer model and with opposite sign than the classical phonon drag known from bulk systems.} Our findings show that, firstly, the thermopower gives important insight into the transport properties of atomic-size structures and second that the linear approximation of the Landauer model has to be used with caution when studying more complex transport properties \ES{even for atomic contacts from free-electron metals.}  
\vspace{0.3cm}
\end{abstract}

\maketitle


Understanding the interplay between electronic and phononic transport in nanoscale conductors is of fundamental interest and important for the development of functional devices, in particular for those combining charge and heat transport.
In bulk systems the behavior is well known and the theoretical models are established \cite{Alam2013}. Also the net charge transport down to the ultimate small size scale of few-atom contacts is meanwhile well understood, in particular in the quantum coherent regime \cite{Agrait2003}.
In this situation, the properties can be described by the Landauer formalism as a wave-scattering process, with an
energy-dependent transmission $\tau(E)$ describing the probability of an electron
to travel from the occupied states on one side of a contact to empty states on the other.
For sufficiently small bias voltage, the current is proportional to the \MS{applied } voltage,
in which case the energy dependence can be neglected and the conductance is given by the simple formula $G = G_0 \tau (E_F)$, with the conductance quantum $G_0 = 2e^2/h$ and $\tau(E_F)$ the transmission probability at the Fermi energy $E_\text{F}$. In atomic contacts of metals \MS{such as } gold, this expression 
\MS{remains a very good approximation even } at room temperature, since the thermal energy $k_BT$ is much smaller than $E_F$.
The thermopower $S$ describes the negative ratio between the voltage measured
across a contact and the temperature difference which creates that voltage.
If the two sides of an atomic contact are at
different temperatures, the Fermi distribution has a softer edge on the hot side
than on the cold side, causing current to flow from the occupied states to the empty states generated by the temperature difference. Using the same approximation as 
\MS{in the derivation of the conductance expression and assuming }
that the temperature difference is small
compared to the absolute temperature $T$, one finds
\begin{equation}S = - \frac{\pi^2 k_\text{B}^2 T}{3 e}
\left(\frac{\partial \ln(\tau(E))}{\partial E}\right)_{E=E_\text{F}}
\label{eq:landauer}
\end{equation}

\noindent \cite{Ludoph1999,Agrait2003}.
This approximation predicts that $S$
depends linearly on  $T$
if $\tau(E)$ is temperature independent. In atomic contacts of gold, $\tau(E)$ is dominated by the $s$-orbitals and is very close to 1, 
\MS{with only a weak energy dependence }
\cite{Cuevas2007}. As a result, one would expect $S$ to be very small, with either negative or positive sign, depending on  the sign of the logarithmic derivative in Eq. \eqref{eq:landauer}.

\begin{figure}
  \includegraphics[width=\columnwidth]{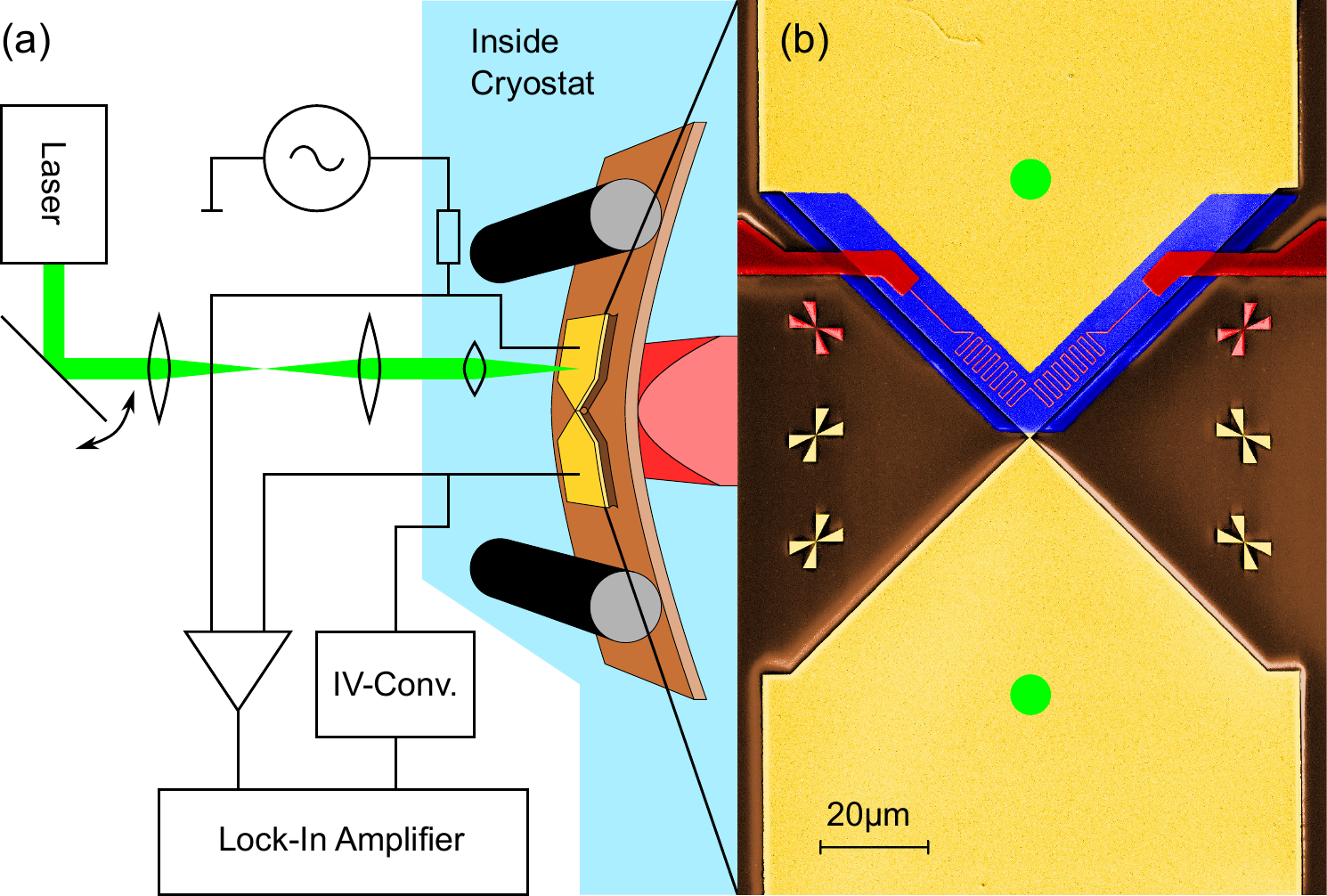}
  \caption{
    (a)
    Scheme of the experimental setup. The sample is bent by retracting the countersupports (black) of a three-point bending mechanism with a fixed central rod (red). The laser beam is steered by $4f$-optics and focused onto the sample (metal structures depicted in yellow, substrate in brown).
    (b) A  colored scanning electron micrograph of the sample shows the constriction made of gold (yellow)
    as well as the local thermometer (platinum resistor with gold leads, red) electrcally isolated from the junction by a
 layer of aluminum oxide (blue). 
    The positions of the laser focus for creating the temperature difference are marked in green.
    \label{fig:setup}
    }
    \end{figure}
In a seminal experiment, Ludoph \emph{et al.} \cite{Ludoph1999} measured the thermopower of gold atomic contacts at an average temperature $\overline{T} = (T_{\mathrm{hot}} + T_{\mathrm{cold}})/2)$ of \SI{12}{\kelvin}, 
\MS{using the MCBJ technique with a notched bulk wire to form the contact. }
Using heaters and thermometers on the wire on both sides of the constriction,
they calibrated the temperature 
\MS{via } the bulk $S$.
Their data shows negligible $S$ when statistically averaging over many different contacts, in agreement with the expectation outlined above. However, they find a large statistical spread, which they attribute to quantum interference effects arising from backscattering of the transmitted electrons at the banks of the contact \cite{Ludoph2000}.
The vanishing $S$ at low temperature was later supported by first-principle calculations \cite{Pauly2011}.
Measurements of $S$ at room temperature were performed by Tsutsui \emph{et al.}
\cite{Tsutsui2013} using a lithographically fabricated MCBJ
with a platinum heater integrated into the lithographic structure.
For a single-atom contact, they found values for $S$
between -2 and -\SI{9}{\micro\volt\per\kelvin}.
Later, they showed that the contact exhibits peculiar behavior when heated to \SI{440}{\kelvin}
\cite{Morikawa2014}.
Matsushita \emph{et al.} \cite{Matsushita2015}\MS{, in contrast, employed } 
a notched-wire MCBJ with thermometers and heaters
on both sides to measure at average temperatures above room temperature and found values
between 0 and \SI{4}{\micro\volt\per\kelvin}.
Evangeli \emph{et al.} \cite{Evangeli2015} measured the thermopower of gold atomic contacts realized in a scanning tunneling microscope at room temperature with the tip heated to generate a temperature difference of \SI{20}{\kelvin} or \SI{40}{\kelvin}.
\MS{They found } an ensemble-averaged $S$ of \SI{-0.75}{\micro\volt\per\kelvin} for contacts with a conductance of up to roughly $100\,\mathrm{G}_0$, 
increasing \MS{toward } the bulk value of $+1.94\,\mu$V/K  \cite{Cusack1958} for larger contacts.
They were able to verify the observed size dependence with theoretical calculations, however, only when going beyond the simplifications included in Eq. \eqref{eq:landauer}, \MS{yielding}
\begin{equation}
 S = -\frac{K_1(T)}{eTK_0(T)}
 \label{eq:fulltheory}
 \end{equation}
taking into account the functions $K_n(T)= \int (E-E_F)^n\tau(E,T)[-\frac{\partial f(E,T)}{\partial E}]\,\mathrm{dE}$ with the Fermi function $f(E,T)$.
Ofarim \emph{et al.} \cite{Ofarim2016} and later Kopp \cite{Kopp2016} 
\MS{measured } the thermopower in an optically heated lithographic gold MCBJ at the fixed temperature \SI{77}{\kelvin}
and found values between -1.5 and \SI{1.5}{\micro\volt\per\kelvin},
with an average clearly below zero, which would be in agreement with a linear interpolation between the low temperature and room temperature values observed in the earlier experiments mentioned above.
In order to verify the partially conflicting experimental findings and to further test the validity of the approximated Eq. \eqref{eq:landauer} for a testbed system, it is therefore insightful to measure the temperature dependence $S(T)$ of gold atomic contacts in the range from $\approx 10\,{\mathrm K}$ to room temperature with the same experimental realization of the contacts.


In the present work the atomic contacts are realized by lithographically fabricated MCBJs.
A \SI{80}{\nano\meter} thick thermally evaporated gold film is patterned by electron beam lithography
on top of a bendable substrate to form a constriction of roughly 100 nm width and is subsequently locally released from the substrate by dry etching to create a short free-standing
bridge around the constriction.
By bending the substrate, this bridge is stretched
\MS{and thereby plastically deformed, thinning down } to an atomic contact (Fig.~\ref{fig:setup}(a)).

Bending the substrate further leads to \MS{the } breaking the bridge and 
\MS{the formation of } a tunnel contact.
When the bending is released, the two ends come back into gentle contact, forming
again an atomic-size contact, but with different atomic arrangement.
The state of the sample is monitored by recording the conductance
as a function of bending in so-called opening and closing traces
(Fig.~\ref{fig:setup}(b)).
A lock-in amplifier provides an ac bias voltage of \SI{0.5}{\milli\volt} which is applied
via a series resistance to the atomic contact.
The current flowing across the contact is detected by a current-voltage converter
with a transimpedance of \SI{1}{\mega\volt\per\ampere}.
To eliminate the effects of the wiring resistance, the voltage drop across
the sample is measured with a differential amplifier in a four-wire resistance measurement scheme.

\begin{figure}[h!]
  \includegraphics[width=0.95\columnwidth]{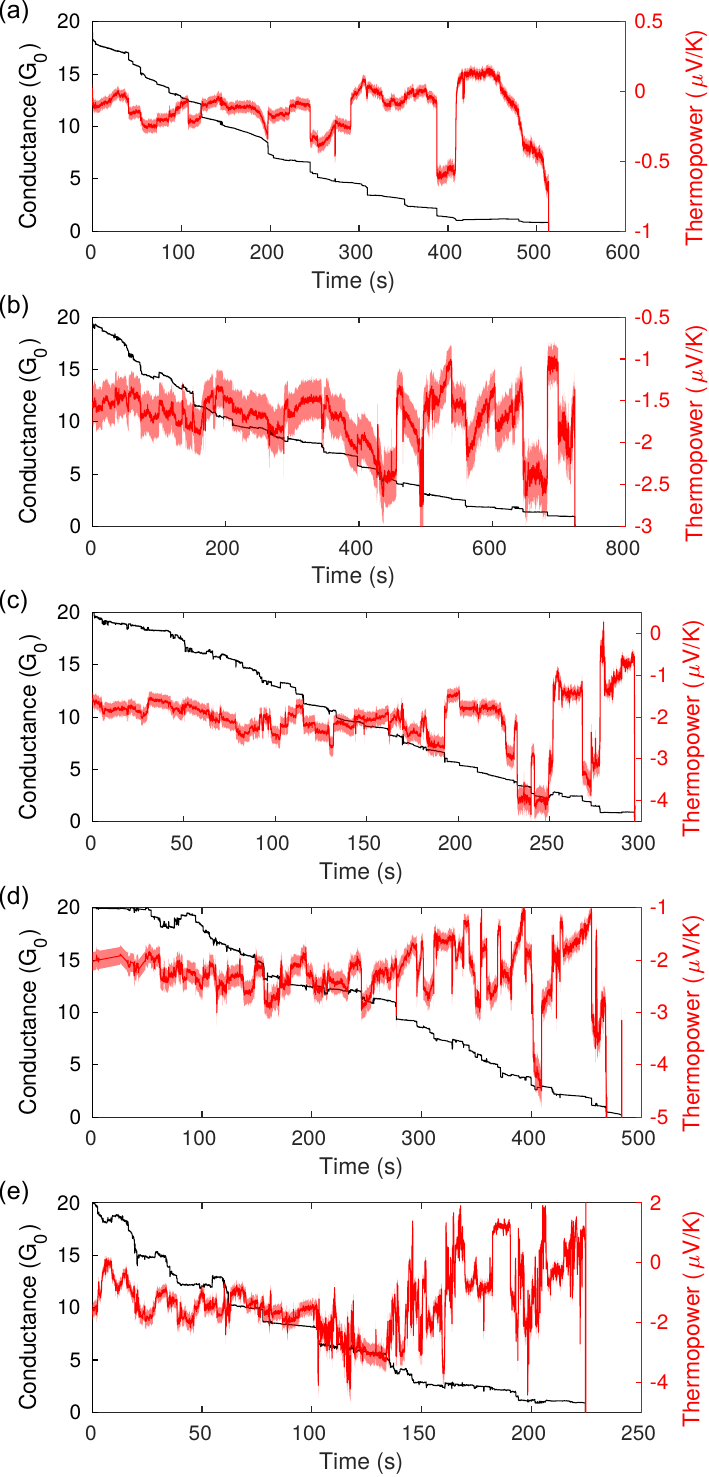}
  \caption{
       The conductance $G$ (black, left axis) and the thermopower $S$ (red, right axis)
    are measured simultaneously during the opening of a nanobridge. The pale frames around the data traces correspond to the standard deviation.
    The measurements were performed at 
   a) $\overline{T} = \SI{18.1}{\kelvin}$, $\Delta T = \SI{12.3}{\kelvin}$;
b)    $\overline{T} = \SI{89}{\kelvin}$ , $\Delta T = \SI{4.3}{\kelvin}$;
  c)   $\overline{T} = \SI{142}{\kelvin}$, $\Delta T = \SI{3.6}{\kelvin}$;
    d) $\overline{T} = \SI{163}{\kelvin}$, $\Delta T = \SI{3.1}{\kelvin}$; 
e) $\overline{T} = \SI{297}{\kelvin}$, $\Delta T = \SI{3.7}{\kelvin}$.
      }
  \label{fig:GandS}
\end{figure}

Since in a MCBJ both sides of the contact reside on the same substrate,
the necessary temperature difference between the two sides has to be created
by local heating.
The latter is performed similar to the approach of Ofarim \emph{et al.} \cite{Ofarim2016}
by illuminating the metal film with a focused laser beam (wavelength \SI{520}{\nano\meter})
at a position about \SI{50}{\micro\meter} away from the atomic contact.
The laser focus is smaller than \SI{10}{\micro\meter} and the power is less than \SI{10}{\milli\watt} generating a temperature difference of several Kelvin at the constriction.
The optical heating also forces the use of a modified mechanical bending mechanism,
where the counter supports are moved instead of the central rod on the rear side of the sample
in order to maintain a constant distance between the sample and the lens focussing the laser.
A simplified schematic drawing of the setup is shown in Fig.~\ref{fig:setup}(a).
The distance to the constriction is 
\MS{sufficiently large } that small errors in the positioning
and the beam shape of the laser focus can be neglected.
It also allows 
\MS{for } a thin resistive temperature sensor,
realized as a platinum meander structure, \MS{to be placed } between the heat source
and the atomic contact.
As shown in Fig.~\ref{fig:setup}(b),
the sensor is lithographically fabricated on top of the gold structures
separated only by an isolating layer of aluminum oxide for electrical isolation.
The resistance of this platinum structure is measured in a four-wire configuration.
The temperature dependence of the resistance is calibrated
with the thermometer in the sample chamber 
\MS{prior to } the thermopower measurement.
The integrated platinum resistor is similar to the one used in Refs.
\cite{Tsutsui2013,Morikawa2014}; 
\MS{however, in our implementation the structure is placed on top of the Au layer to minimize heat flow into the substrate, and is used solely for the temperature measurement, since local heating is provided optically.}

During the thermopower measurement procedure,
the atomic contact is \MS{repeatedly } opened and closed while the laser 
\MS{continuously heats } the gold structure, at variance to Ref. \cite{Ofarim2016}
\MS{, where a pulsed laser was used. This approach, however, led to thermal expansion that destabilized the atomic contact and complicated the analysis. }
On every second opening-closing trace,
the laser is turned off such that the 
\MS{subsequently } recorded trace can serve as a reference.
To exclude thermovoltage generated by the wiring,
the heating spot is moved from one side of the contact
to the other for every second opening-closing trace.
Hence, traces with and without laser heating are recorded
with the laser \MS{positioned } on either side of the constriction.

The ac components of the current and voltage across the contact are
used to calculate the conductance of the atomic contact.
The generated thermovoltage $V_{\mathrm{th,meas}}$ is evaluated 
\MS{from } the dc components.
$V_{\mathrm{th,meas}}$ also contributes to the measured current,
since the biasing circuit shunts the generated $V_{\mathrm{th,meas}}$,
but the open-circuit thermovoltage $V_\mathrm{th}$ can be calculated \MS{by } knowing the impedance of the biasing circuit.
Since any drift in the amplifier offsets appears as voltage signal as well,
the offsets are subtracted by using the opening-closing trace recorded
without heating just before the measured trace.

The temperature difference across the sample is determined
from the resistance of the platinum sensor
when the laser heats the side with the platinum structure, 
\MS{compared to the value when the laser heats the opposite side; for details, see the Supporting Information (SI). } In the SI we also describe finite-element simulations of the static heat flow, which show that the temperature increase at the constriction is about \SIrange{75}{80}{\percent}
of the temperature increase at the Pt sensor, due to the sample geometry.
This correction was applied to the measured temperature difference (see SI for more details).

Fig.~\ref{fig:GandS} shows typical results for several opening traces recorded at five different temperatures.
During the bending of the substrate, the atomic configuration of the contact changes
in \MS{discrete } steps due to atomic rearrangements.
These rearrangements are well known \cite{Agrait2003},
and the conductance is mainly determined by the structure of the 
\MS{narrowest } part of the contact.
The bending results in a step-like conductance change, which can be represented
in a conductance histogram as shown in Fig.~S7 in the SI.
This histogram is material specific and agrees well with the known behavior
of gold atomic contacts, revealing a 
\MS{pronounced } maximum around 1~$\mathrm{G_0}$
and several smaller maxima close to integer multiples of $\mathrm{G_0}$ \cite{Agrait2003,Ludoph2000}.


%

\begin{figure}
  \includegraphics[width=1.0\columnwidth]{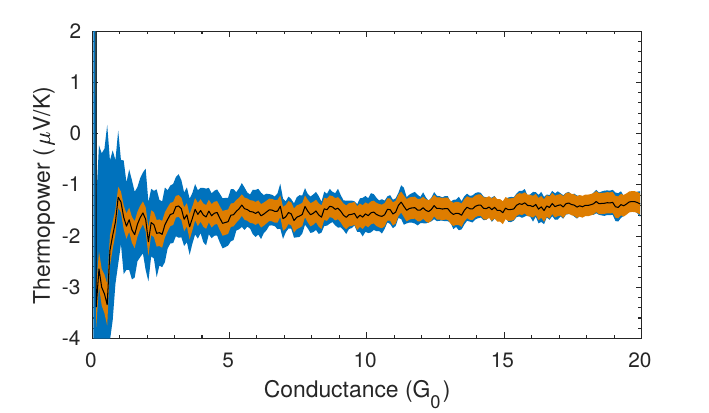}
  \caption{The measured thermopower is collected in bins of \SI{0.1}{G_0} and averaged.
    This average is shown as function of the conductance (black line).
    The blue area indicates the standard deviation of $\langle S\rangle$.
    The measurement uncertainty of the temperature difference between the hot and the cold side
    is propagated to the uncertainty in the thermopower indicated by the orange area.
    The data contains 48 opening traces and was recorded at $\overline{T} = \SI{91}{\kelvin}$.
   }
    \label{fig:SoverG}
\end{figure}

The thermopower varies much less than the conductance over 
\MS{single } opening trace. Instead, there are small jumps towards \MS{both } higher as well as towards lower $S$, \MS{closely } coinciding with the conductance steps. Fig.~\ref{fig:GandS} shows that the average thermopower is negative and \MS{reaches its } lowest \MS{value } at intermediate temperatures.  In addition,  $S$ is influenced significantly by quantum interference of electronic paths being back reflected at the electrodes \cite{Agrait2003,Cuevas2007,Ludoph2000}.
This means that $S$ may change even if the central contact region remains
unaffected by the bending.
This can also be seen in Fig.~\ref{fig:GandS}, where $S$ shows pronounced variations
in regions where $G$ reveals featureless plateaus, in agreement with \cite{Ludoph2000}.

The influence of the banks is random and increases with decreasing conductance,
resulting in fluctuations around a mean value $\langle S\rangle$ with increasing amplitude for decreasing $G$
\cite{Ludoph1999,Evangeli2015,Matsushita2015}.
To visualize these fluctuations, the thermopower of many traces was averaged and is shown in
Fig.~\ref{fig:SoverG}, with the blue area indicating the standard deviation.
The measurement uncertainty of the temperature difference between the hot and the cold sides
is propagated to the uncertainty in the thermopower, indicated by the orange area.
In agreement with the aforementioned reports, $\overline S$ shows no pronounced
dependence on the conductance in the range from \SIrange{1}{20}{G_0},
justifying the averaging of $\langle S\rangle$ over this conductance range,
resulting in $\overline{\langle S\rangle}$.  We do not observe the pronounced minima of the fluctuations around $G = 1\,\mathrm{G_0}$ reported earlier \cite{Ludoph1999,Evangeli2015,Matsushita2015}, although the conductance histogram clearly shows the high abundance of such contacts, which are identified as single-atom contacts or chains. We also note that for  $G \sim 1G_0$, the absolute size $|\overline{\langle S\rangle}|$ is smallest. For conductance smaller than 0.7\,$\mathrm{G_0}$, $\langle S\rangle$ is strongly negative, in agreement with the theoretical calculations of Evangeli \emph{et al.}, \cite{Evangeli2015}, which were, however, not 
\MS{reproduced } in their (room temperature) measurements. Nevertheless this observation supports the applicability of the simulation model also at this intermediate temperature \MS{regime}.

\begin{figure}
  \includegraphics[width=1.0\columnwidth]{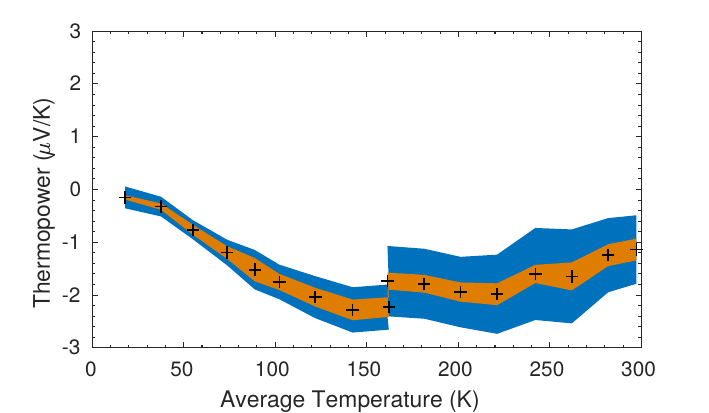}
  \caption{The averaged thermopower $\overline{\langle S\rangle}$ is shown as a function
    of the average temperature $\overline T$ of the contact (black symbols).
    The results show a clear nonmonotonic behavior with a pronounced minimum around
    \SI{150}{\kelvin}.
    The discontinuity at about \SI{160}{\kelvin} is due to the fact that the sample
    had to be replaced at that temperature.
    The blue area indicates the standard deviation of the thermopower measurements
    due to the fluctuations in the contact geometry and the orange area indicates
    the uncertainty resulting from the temperature measurement.}
  \label{fig:SoverT}
\end{figure}

The installation of the setup in a variable-temperature cryostat enables us to perform
these measurements at different temperatures.
The mean thermopower $\overline{\langle S\rangle}$ is shown in Fig.~\ref{fig:SoverT} as function
of the average temperature $\overline T$ of the atomic contact.
The data reveals a continuous dependence of $\overline{\langle S\rangle}$ on $\overline T$
starting at about \SI{0}{\micro\volt\per\kelvin} at the low-temperature end,
in agreement with the results of Ref. \cite{Ludoph1999},
and 
 \MS{reaching } -\SI{1.1}{\micro\volt\per\kelvin} at room temperature.
Evangeli \emph{et al.} \cite{Evangeli2015} found -\SI{0.75}{\micro\volt\per\kelvin}
at an average temperature in the range of \SIrange{305}{315}{\kelvin},
which agrees with an extrapolation of our data as well.

\begin{figure*}[t]
    \centering
    \includegraphics[width=1.65\columnwidth]{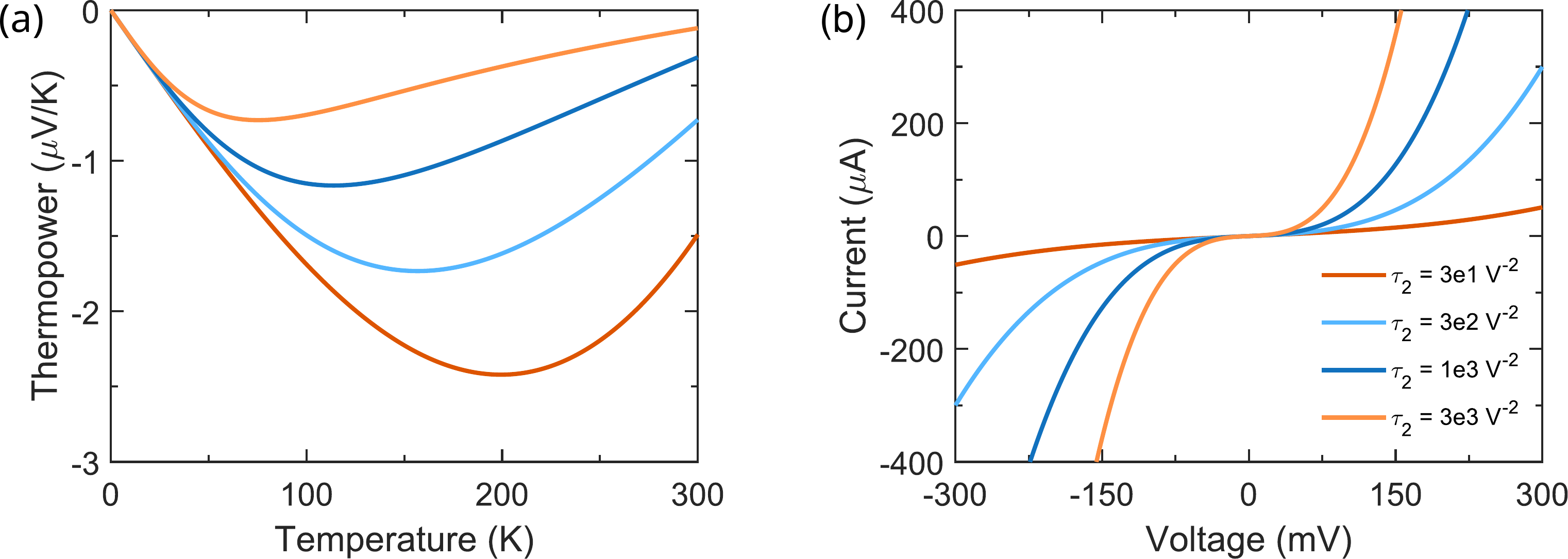}
    \caption{\MS{Model-based higher-order corrections to the thermopower of purely electronic origin: Calculated nonmonotonic evolution of the thermopower (a) and corresponding $IV$ characteristics (b) for varying values of $\tau_2$, while $\tau_0=1$, $\tau_1=-0.76$\,V$^{-1}$ and $\tau_3=344.3$\,V$^{-3}$ are kept constant.}}
    \label{fig:thermopower_model}
\end{figure*}

Interestingly, $\overline{\langle S\rangle}$ shows a pronounced minimum
between \SI{100}{\kelvin} and \SI{200}{\kelvin}.
Note that at \SI{160}{\kelvin} the sample had to be exchanged, resulting in two data points for this temperature and reflecting the uncertainty in the determination of the absolute value of $S$. Possible origins for the uncertainty are variations in the geometry of the freestanding bridge. The calibration was performed for one sample, and also in the simulation an idealized geometry was assumed. As the simulations also show, the thermal conductance of the atomic-scale constriction is negligible. However, the thermal coupling of the electrodes depends, e.g., on the suspended length, which slightly varies from sample to sample. These differences give rise to variations in the actual temperature difference at the atomic constriction 
\MS{compared to the simulated one. }
However, the decreasing trend of $\overline{\langle S\rangle}$ between \SI{20}{\kelvin}
and \SI{140}{\kelvin},
and the increasing behavior between \SI{160}{\kelvin} and room temperature, are statistically
significant and not affected by these sample-to-sample variations.
The nonmonotonic dependence was confirmed
over the complete temperature range using another sample
and slightly different measurement electronics, \MS{albeit with } somewhat smaller statistics
and less stable measurement conditions. The results are shown in Fig.~S8
in the SI.

To interpret this finding, we first note that in our experiment, the thermopower is negative over the entire investigated temperature range, as expected within the free-electron model for electron transport. The positive value of bulk Au reported in literature \cite{Cusack1958} arises from Fermi surface effects beyond the free-electron model \cite{Mahan2016}. The negative value observed here thus indicates that these band-structure properties are negligible in atomic contacts and \MS{that } the free-electron model is more appropriate. A nonmonotonic temperature dependence of $S$ is, however, in
disagreement with the linearized Landauer model of Eq. \eqref{eq:landauer},
which predicts a linear temperature dependence of $S$,
if the energy dependence of the transmission is independent of temperature. We are not aware of a theoretical treatment of the temperature dependence of $S$ for Au atomic contacts. However, for dimer structures of $\mathrm{C}_{60}$ between gold electrodes, a stronger-than-linear, but still monotonic temperature dependence has been calculated \cite{Kloeckner2017}. Similarly, the thermopower of Au-benzenedithiol-benzene single-molecule junctions has been observed experimentally to increase linearly with temperature in the range of \SIrange{100}{320}{K} \cite{Kim2016}.\\

\MS{We firstly discuss a theoretical model that provides a possible explanation for a nonmonotonic thermopower, that is capable of reproducing a similar nonmonotonic trend in the thermopower as observed in our experiment. As the model assumes a purely electronic origin, which should simultaneously manifest in the current-voltage characteristics ($IV$s)
\begin{equation}
I(V) = \int \tau(E)\left[f_L(E)-f_R(E)\right]\mathrm dE;,
\end{equation}
this interpretation opens the door for further experimental verification via ensemble-averaged transport measurements on atomic contacts.}

As pointed out by Evangeli \emph{et al.} \cite{Evangeli2015}, Eq. \eqref{eq:landauer} was insufficient to reproduce the experimentally observed magnitude, sign and size dependence of the room temperature thermopower of Au atomic contacts. Instead the more complex theory expressed by Eq. \eqref{eq:fulltheory} 
\MS{must } be applied. Eq. \eqref{eq:fulltheory} makes the physical meaning of $S$ apparent: it is a measure for the energy content of the current (per charge) and temperature and thus corresponds to the entropy related to the current (per charge). An energy content 
\MS{that increases quadratically with } temperature, as it is the case, e.g., for the Sommerfeld model, results in a linear temperature dependence of $S$. Any deviation from the Sommerfeld free-electron model can, in principle, break the linear temperature dependence. This \MS{would }require a peculiar temperature and energy dependence of $K_1$ and $K_0$ to produce a nonmonotonic $S(T)$ within this approach. 

\MS{

\begin{figure*}[t]
    \centering
    \includegraphics[width=1.7\columnwidth]{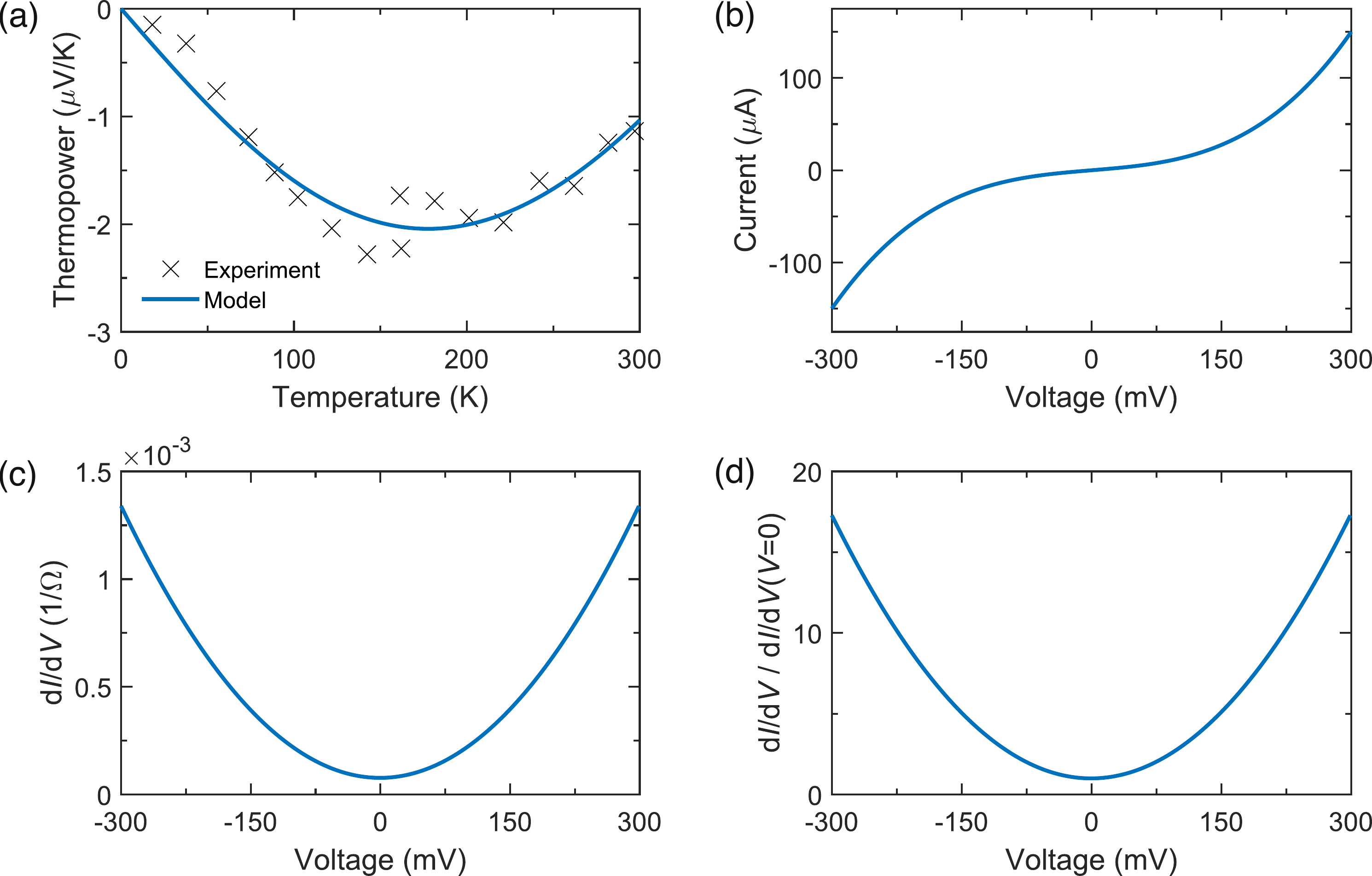}
    \caption{\MS{(a) Experimental data of the temperature-dependent thermopower (marker symbols) and fit to the model (solid line) using Eq. \eqref{eq:thermopower_model}. The fit parameters $\tau_0$, $\tau_1$, $\tau_2$ and $\tau_3$ serve as parameterization for the calculation of the (b) $IV$ characteristics, (c) differential conductance d$I$/d$V$, and (d) normalized differential conductance (d$I$/d$V$)/(d$I$/d$V$($V=0$)), following Eq. \eqref{eq:IV_thermo_model}.}}
    \label{fig:thermopower_fit}
\end{figure*}

Generally speaking, the approach incorporates higher-order corrections to the quantum thermopower by considering an energy-dependent transmission function $\tau(E)$, which can be expanded as
\begin{equation}
\tau(E) = \sum_{n=0}^{\infty} \frac{1}{n!} \tau_n E^n = \tau_0 + \tau_1 E + \frac{1}{2} \tau_2 E^2 + \frac{1}{6} \tau_3 E^3\;,
\end{equation}
where $\tau_n = \left( \partial^n \tau(E)/\partial E^n \right)|_{E=0}.$ A further expansion of the voltage-dependent term $f_L(E) - f_R(E)$ reveals an even symmetry in energy, implying that only $\tau_0$ and $\tau_2$ contribute to the $IV$. As a result, the current can then be expressed as
\begin{equation}\label{eq:IV_thermo_model}
    I(V,T) = I_1(T) V + I_3(T) V^3
\end{equation}
with the two contributions
\begin{align}
    I_1(T) &= \tau_0 + \tau_2 \dfrac{2\pi^2}{3} T^2 = \tau_0 + \tau_2 \beta T^2 \\
    I_3(T) &= \tau_2 \dfrac{1}{6} \left( \dfrac{\pi^2}{6} + 1 \right) = \alpha \tau_2
\end{align}

Similarly, thermal transport can be evaluated by calculating the transport coefficient
\begin{align}
    K_1 &= \int E\,\tau(E) \left( - \frac{\partial f}{\partial E} \right) \text{d}E \\
    &\overset{\text{Exp.}}{=}\;\;\; \dfrac{\pi^2}{3} T^2 \left( \tau_1 + \dfrac{7\pi^2}{30} T^2 \tau_3 \right)\;.
\end{align}
Combining this result with the expression for the conductance, where $G = \text{d}I/\text{d}V|_{V=0} = I_1$, and introducing the constant $\alpha' = 7\pi^2/30$, one obtains the Seebeck coefficient 
\begin{equation}\label{eq:thermopower_model}
    S = \dfrac{K_1}{TG} = \dfrac{\pi^2 T}{3} \dfrac{\tau_1 + \alpha' \tau_3 T^2}{\tau_0 + \beta \tau_2 T^2}\;.
\end{equation}
With this expression for $S(T)$, the thermopower shows indeed a nonlinear dependence on the temperature, with the precise analytical trend being governed by all four coefficients $\tau_0$, $\tau_1$, $\tau_2$ and $\tau_3$. At the same time, a similarly related behavior should also be reflected in the nonlinear conductance, given by
\begin{equation}
    G = \dfrac{\text{d}I}{\text{dV}} = I_1 (T) + 3 I_3 (T) V^2\;.
\end{equation}

\noindent The model’s predictions are visualized in Fig. \ref{fig:thermopower_model}(a), where $S(T)$ is calculated for varying values of $\tau_2$, while keeping the other parameters in Eq. \eqref{eq:thermopower_model} constant. With the variation of the (even) coefficient $\tau_2$ in Eq. \eqref{eq:IV_thermo_model}, corresponding changes in the nonlinear $I$–$V$ characteristics are shown in Fig. \ref{fig:thermopower_model}(b).

\begin{figure*}[t]
    \centering
    \includegraphics[width=1.8\columnwidth]{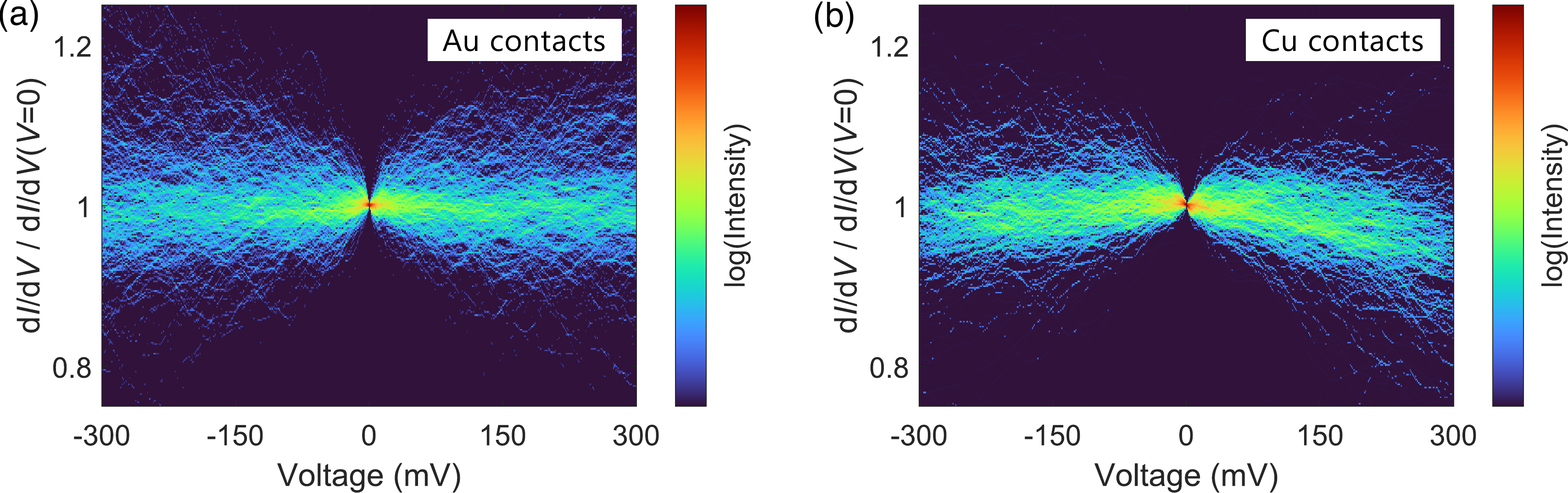}
    \caption{\MS{2D histograms of the normalized differential conductance (d$I$/d$V$)/(d$I$/d$V$($V=0$)). (a) Au atomic contacts: Ensemble of 452 individual measurements with zero-bias conductance d$I$/d$V$($V=0$) between 0.6 and 2\,G$_0$. (b) Cu atomic contacts: Ensemble of 340 individual measurements with d$I$/d$V$($V=0$) between 0.6 and 6\,G$_0$.}}
    \label{fig:thermopower_dIdV}
\end{figure*}

In the next step, Eq. \eqref{eq:thermopower_model} shall be used to analyze the nonmonotonic thermopower of Au atomic-size contacts. A fit to the experimental data, which is presented in Fig. \ref{fig:thermopower_fit}(a), highlights the appeal of the model. We obtain a good agreement with the experiment over the entire temperature range, considering that both the low-temperature and room temperature values even align with the literature reports on atomic-scale contacts. With the knowledge of the fitparameter $\tau_0$, $\tau_1$, $\tau_2$ and $\tau_3$, the coefficients can be partially used to parametrize the $I$–$V$ curve, as shown in \ref{fig:thermopower_fit} (b). Further the differential conductance d$I$/d$V$ and the normalized differential conductance (d$I$/d$V$)/(d$I$/d$V$($V=0$)) are computed in Fig. \ref{fig:thermopower_fit}(c) and (d), respectively. In particular, the latter provides a concrete benchmark for a comparisons with transport measurements. According to the model, the nonmonotonic temperature dependence of the thermopower translates into an upward-parabolic d$I$/d$V$ curve. Specifically, taking 300\,mV as a reference, the differential conductance increases by a factor of about 15 relative to its zero-bias value.}

Such a functional dependence would be expected to be very sensitive to the detailed atomic configuration and is expected to average out when studying a large ensemble of contacts. However, we observe the nonmonotonic behavior in the ensemble-averaged data set. 
Such a functional dependence of $\tau(E)$ should also be reflected in the $IV$s of atomic contacts on the voltage scale $eV \sim 3.5 k_{\mathrm B}T$, which corresponds to approximately 90 meV at room temperature. To elucidate this option, we analyze the $IV$s of a few hundred atomic contacts in the conductance range of \SIrange{1}{7}{G_0} recorded at low temperature ($\leq 4.2\,$K). For larger contacts, the analysis is hampered by \MS{discontinuous conductance } jumps 
\MS{driven by } atomic rearrangements \cite{Schirm2013,Ring2020,Strohmeier2025}. 

\MS{On a similar voltage range, transport measurements at 4\,K were carried out on Au and Cu MCBJ. Both materials serve as a suitable testbed as they form monometallic atomic contacts. Fig. \ref{fig:thermopower_dIdV} presents 2D histograms summarizing ensembles of d$I$/d$V$ measurements recorded on a variety of different atomic contacts. As a data basis, configurations with a zero-bias conductance between 0.6 and 2\,G$_0$ for Au, and 0.6 and 2\,G$_0$ for Cu are considered. From a closer look on the colour-coded intensity, its is evident that within a few hundred mV, deviations in the differential conductance remain relatively small compared to the parabolic trend predicted by the model approach. The maximum variation is estimated to be around 20\,\%. For Cu, even a downward tend in the normalized differential conductance (d$I$/d$V$)/(d$I$/d$V$($V=0$)) appears. Given this mismatch, we cannot verify the nonlinear transport properties predicted as a consequence of higher-order corrections to the thermopower. Consequently, the overall validity of the model remains in question. It seems plausible that instead phonon contributions or other degrees of freedom play a crucial role.

}



We therefore argue that the minimum arises from non-electronic contributions to $S$ caused by the interaction with the lattice. This assumption is motivated by the observation that the minimum occurs at $\sim$150\,K, which is comparable to the bulk Debye temperature 
\MS{of gold, reported as 165\,K }
\cite{Kittel2005}. For clean bulk metals, the phonon-drag effect, which results from moving electrons emitting phonons
\MS{that co-propagate }
with the electron and thereby contributing to the energy transport, gives rise to a {\em maximum} of $S$ well below the Debye temperature \cite{Cusack1958}. This phonon-drag contribution is positive, \ES{contrary to our observations}. In dirty systems, the peak vanishes because the phonons may loose their energy by decaying into other phonons \cite{Mahan2016}.

 However, also for negative $S$ the phonon drag should give rise to a positive contribution to the thermopower, 
 \MS{rather than a dip as observed here. } For clean bulk Au, the phonon-drag peak occurs around 40 to 70 K \cite{Cusack1958}, i.e., in a temperature range where we do observe a monotonic decay. We thus argue that firstly the phonon-drag effect in atomic contacts is suppressed by the filtering effect of the atomic constriction which greatly reduces phonon transport. \ES{A  theoretical model that reproduces this intriguing observations is currently lacking and beyond the scope of this article.} 
 
 
 Other possible explanations include structural effects, which could change the geometry of the preferred 
 \MS{atomic configurations } and thereby the transmission coefficients. Although we cannot rule out such a possibility, 
 \MS{we also do not have } indications of such an effect.

In conclusion, we report a \ES{combined experimental and theoretical} study of the temperature dependence of the thermopwer of atomic-size contacts of Au using nanofabricated break junctions, optical heating for creating the temperature gradient and a local temperature measurement. We observe a nonmonotonic temperature dependence, which, in the limiting cases of low and room temperature, agrees with the values reported in literature. In between, the thermopower reveals a minimum that cannot be described \ES{by purely electronic properties, even when going beyond } the linear approximation of the Landauer theory, which predicts a negligibly small thermopower at low temperature and a linear increase in magnitude with the absolute temperature. Possible origins of the observed complex temperature dependence are either 
\ES{a result of unusual electron-phonon coupling, that yields a 'phonon repel' effect, } 
or structural changes. Hence, our study reveals that the thermopower, even in a testbed system for quantum transport such as Au atomic contacts, behaves more complex than expected. It thus underlines that 
\MS{thermopower measurements } give insight into the transport properties well beyond what can be revealed by the electronic conductance \MS{alone}.

\section*{Acknowledgment}
\ES{We want to thank J. C. Cuevas, P. Nielaba and C. B. Winkelmann for fruitful discussions and D. Majidi, K. Kirchberger and R. Zerfass for their contributions in the early phase of the project.
We gratefully acknowledge the financial support of the
Deutsche Forschungsgemeinschaft (DFG) via the Collaborative Research Center SFB767 and project 493158779 as part of the Spezialforschungsbereich (SFB) Q-M\&S of the Austrian Science Fund (FWF, project DOI 10.55776/F86)}
The measurements would also not have been possible without the support of the nano.lab as well as the machine and electronics workshops of the university.

\bibliography{literature}
\end{document}